\documentclass[aps,pre,superscriptaddress,preprint]{revtex4-1}%
\usepackage{amsfonts}
\usepackage{amsmath}
\usepackage{amssymb}
\usepackage{graphicx}%

\begin{document}
\preprint{ }
\title{Circularly polarized modes in magnetized spin plasmas}

\author{A. P. MISRA}
\affiliation{Department of Physics, Ume\aa \ University, SE-901 87 Ume\aa , Sweden}

\author{G. BRODIN}
\affiliation{Department of Physics, Ume\aa \ University, SE-901 87 Ume\aa , Sweden}

\author{M. MARKLUND}
\affiliation{Department of Physics, Ume\aa \ University, SE-901 87 Ume\aa , Sweden}

\author{P. K. SHUKLA}
\altaffiliation[Also at: ]{Institut f\"{u}r Theoretische Physik IV, Fakult\"{a}t f\"{u}r Physik and
Astronomie, Ruhr-Universit\"{a}t Bochum, D-44780 Bochum, Germany}
\affiliation{Department of Physics, Ume\aa \ University, SE-901 87 Ume\aa , Sweden}

\begin{abstract}
The influence of the intrinsic spin of electrons on the propagation of
circularly polarized waves in a magnetized plasma is considered. New
eigenmodes are identified, one of which propagates below the electron
cyclotron frequency, one above the spin-precession frequency, and another
close to the spin-precession frequency.\ The latter corresponds to the spin
modes in ferromagnets under certain conditions. In the nonrelativistic motion
of electrons, the spin effects become noticeable even when the external
magnetic field $B_{0}$ is below the quantum critical\ magnetic field strength,
i.e., $B_{0}<$ $B_{Q} =4.4138\times10^{9}\, \mathrm{T}$ and the electron density
satisfies $n_{0} \gg n_{c}\simeq10^{32}$m$^{-3}$. The importance of \ electron
spin (paramagnetic) resonance (ESR) for plasma diagnostics is discussed.

\end{abstract}
\maketitle

\section{Introduction}

During recent years there has been a rapid increase in the interest of quantum
plasmas, see e.g., Refs.
\cite{Padma-Nature,Ferro-PRE,Marklund-Trieste,Tito-2008a,Tito-2008b,Tito-Rydberg-plasma,Manfredi-review,Shukla-Eliasson-review,Manfredi-quantum-well,spinMHD,Classical-quant,Brodin2008,spinPRL,Zamanian2010,Asenjo2009}%
. This has been stimulated by experimental progress in nano-scale plasmas
\cite{Manfredi-quantum-well}, ultracold plasmas \cite{Ultracold}, spintronics
\cite{Spintronics}, and plasmonics \cite{Atwater-Plasmonics}. However, already
more than forty years ago, Iannuzzi \cite{spin1} established the possibility
for the existence of electron spin (paramagnetic) resonance (ESR) in a fully
ionized low-temperature plasma, and predicted its importance, e.g., in the
plasma diagnostics for measuring the particle density with a greater
precession than the conventional technique, in determining the particle
velocity spectrum perpendicular to the magnetic field, as well as to calculate
the magnetic field in the propagation of electromagnetic (EM) waves (e.g.,
whistlers, Alfv\'{e}n waves, shock waves etc.) in plasmas. Recent
investigation \cite{EPR} along this line indicates that besides the currently
prevalent laser methods, ESR technique can successfully be used for plasma
diagnostics, e.g., measuring the electron densities in the microwave region.
\ Furthermore, the importance of spin effects in plasmas has also been studied
using kinetic plasma theory
\cite{Brodin2008,Asenjo2009,Zamanian2010,Cowley-1986,Kulsrud-1986}, with
applications to wave propagation \cite{Brodin2008,Asenjo2009,Zamanian2010} as
well as other phenomena \cite{Cowley-1986,Kulsrud-1986}. The hydrodynamic
description of spin plasma waves can also be found in the literature\ (see,
e.g. Refs. \cite{HydrodynamicSpin,Marklund-Trieste,Ferro-PRE}).

Although, whistler waves have been studied for almost a century, they are
still a subject of intense research in view of its importance not only in
space plasmas, but also in many astrophysical environments, e.g., in the
atmosphere of neutron star envelope, in the coherent radio emission in pulsar
magnetosphere etc. Quasilinear theory \cite{quasilinear} and simulation
\cite{simulation}\ show that whistler waves can be used to resonantly
accelerate electrons. \ Furthermore, they can also be used to interpret the
fine structure of Zebra-type patterns and fiber bursts in solar type II and IV
radio bursts \cite{burst}. Thus, the occurrence of ESR might be useful for the
electron acceleration in the propagation of EM radiation in the pulsar
magnetosphere as well as for plasma diagnostics in the microwave region in
laboratory experiments if the conditions favor.

In the present work, we will derive and analyze the dispersion relation for
the propagation of circularly polarized (CP) EM \ (CPEM) waves in a magnetized
spin plasma using a spin fluid model. Various fluid models are appropriate in
different regimes (see e.g., Refs. \cite{Classical-quant,Marklund-Trieste}).
The model to be used contains the Bohm-de Broglie potential, the magnetic
dipole force and includes the spin-precession dynamics as well as the spin
magnetization current. Its basis can be found in e.g., Ref.
\cite{Marklund-Trieste}, and more rigorous foundation can be given starting
from the kinetic theory presented in Ref. \cite{Zamanian2010}. Specifically,
we will focus our discussion to the ESR as well as the properties of spin
modified whistler-like modes.

\section{Weakly nonlinear evolution}

The nonrelativistic evolution of spin$-1/2$ electrons can be described by
\cite{Marklund-Trieste}
\begin{align}
\left(  \partial_{t}+\mathbf{v}_{e}.\nabla\right)  \mathbf{v}_{e}  
=-\frac{e}{m_{e}}\left(  \mathbf{E}+\mathbf{v}_{e}\times\mathbf{B}\right)
-\nabla P_{e}/m_{e}n_{e}  +\frac{\hbar^{2}}{2m_{e}^{2}}\nabla\left(  \frac{\nabla^{2}\sqrt{n_{e}}%
}{\sqrt{n_{e}}}\right)  +\left(  \frac{2\mu}{m_{e}\hbar}\right)
\mathbf{S}.\nabla\mathbf{B}, \label{2}%
\end{align}%
\begin{equation}
\left(  \partial_{t}+\mathbf{v}_{e}.\nabla\right)  \mathbf{S}=-\left(
2\mu/\hbar\right)  \left(  \mathbf{B}\times\mathbf{S}\right)  , \label{3}%
\end{equation}
where $n_{e},$ $m_{e},$ $\mathbf{v}_{e}$ respectively represent the number
density, mass and velocity of electrons, $\mathbf{E}$ $(\mathbf{B})$ is the
electric (magnetic) field, $P_{e}$ is the electron pressure. Also,
$\mathbf{S}$ is the spin angular momentum with its absolute value $\left\vert
\mathbf{S}\right\vert =\hbar/2;$ $\mu=-\left(  g/2\right)  \mu_{B},$where
$g\approx2.0023193$ is the electron $g$-factor and $\mu_{B}\equiv
e\hbar/2m_{e}$ is the Bohr magneton. The equations are then closed by the
Maxwell equations.%
\begin{equation}
\nabla\times\mathbf{E}=-\partial_{t}\mathbf{B},\nabla.\mathbf{B}=0, \label{4}%
\end{equation}%
\begin{equation}
\nabla\times\mathbf{B}=\mu_{0}\left(  \varepsilon_{0}\partial_{t}%
\mathbf{E-}en_{e}\mathbf{v}_{e}\mathbf{+}\left(  2\mu/\hbar\right)
\nabla\times n_{e}\mathbf{S}\right)  , \label{5}%
\end{equation}
The above equations (\ref{2}) and (\ref{3}) apply when different spin states
(i.e., spin-up and spin-down relative to the magnetic field) can be well
represented by a macroscopic average. This may occur for very strong magnetic
fields (or a very low temperature), where generally the lowest energy spin
state is populated. Alternatively, when the dynamics on a time-scale longer
than the spin-flip frequency is considered, the macroscopic spin state is
well-described by the thermodynamic equilibrium spin configuration, and the
above model can still be applied. In the later case, the macroscopic spin
state will be attenuated by a (thermodynamic) factor decreasing the effective
value of $\left\vert \mathbf{S}\right\vert $ below $\hbar/2$. However, this
case will be not considered further in the present paper. As a consequence,
our studies will be focused on the regime of strong magnetic fields, as seen
in astrophysical plasmas \cite{Astroplasma}.

In what follows, we will assume the propagation of a CPEM waves to be of the
form $\mathbf{E}=\left(  \mathbf{\hat{x}}\pm i\mathbf{\hat{y}}\right)
E(z,t)\exp(ikz-i\omega t)+$ c.c., along an external magnetic field
$\mathbf{B}=B_{0}\mathbf{\hat{z}}$, where $E(z,t)$ is the weakly modulated
wave amplitude (i.e. we assume $\left\vert \left(  1/f\right)  \partial
f/\partial z\right\vert \ll k,\left\vert \left(  1/f\right)  \partial f/\partial
t\right\vert \ll \omega$, for all variables $f$), \ $k$ $(\omega)$ represents
the EM wave number (frequency) and c.c. denotes the complex conjugate. In the
interaction of high-frequency (hf) EM waves with the\ hf electron plasma
response, the use of cold plasma approximation is well justified in view of
the fact that for large field intensities and moderate electron temperature,
the directed velocity of electrons in the hf fields is much larger than the
random thermal speed. Moreover, it can also be shown that the density
perturbation associated with the high-frequency (hf) EM wave is zero. \ Thus,
taking the curl of Eq. (\ref{2}) and using Eqs. (\ref{3})-(\ref{5}) we readily
obtain the following evolution equation for CPEM waves.
\begin{align}
0  &  =\frac{e}{m_{e}}\partial_{t}\mathbf{B}+\frac{\varepsilon_{0}}{en_{e}%
}\partial_{t}\left(  \partial_{t}^{2}\mathbf{B}+\frac{1}{n_{e}}\nabla
n_{e}\times\partial_{t}\mathbf{E}\right)  -v_{ez}\nabla\times\partial
{z}\mathbf{v}_{e}+\frac{1}{e\mu_{0}}\partial_{t}\left[  \frac{1}{n_{e}}%
\nabla\times\left(  \nabla\times\mathbf{B}\right)  \right] \nonumber\\
&  -\frac{2\mu}{e\hbar}\partial_{t}\left[  \frac{1}{n_{e}}\nabla\times\left(
\nabla\times n_{e}\mathbf{S}\right)  \right]  +\frac{1}{m_{e}\mu_{0}n_{e}%
}\nabla\times\left[  \left(  \nabla\times\mathbf{B}\right)  \times
\mathbf{B}\right]  +\frac{2\mu}{m_{e}\hbar}\nabla\times\left(  S^{a}\nabla
B_{a}\right) \nonumber\\
&  -\frac{\varepsilon_{0}}{m_{e}n_{e}}\nabla\times\left(  \partial
_{t}\mathbf{E}\times\mathbf{B}\right)  -\frac{2\mu}{m_{e}\hbar n_{e}}%
\nabla\times\left[  \left(  \nabla\times n_{e}\mathbf{S}\right)
\times\mathbf{B}\right]  \label{ev}%
\end{align}
The weakly nonlinear equation (\ref{ev}) is a useful result when considering
the interaction between low-frequency (lf) and hf fields, where the lf fields
are induced by the ponderomotive force. Equation (\ref{ev}) then needs to be
complemented by equations for the lf dynamics, and naturally the number of
dependent variables can be further reduced. However, before this line of
research is pursued, a more through study of the linear theory should be made,
as will be undertaken in the next section.

\section{Linear theory of whistler waves}

Introducing the variables $B_{\pm}=B_{x}\pm iB_{y},$ $E_{\pm}=E_{x}\pm iE_{y}$
etc., suitable for CP waves, and limiting ourselves to the linearized theory,
we obtain respectively from the Faraday's law and the spin-evolution equation
\cite{spinPRL}.%

\begin{equation}
B_{\pm}=\pm\frac{ik}{\omega}E_{\pm},\text{ }S_{\pm}=\mp\frac{2\mu\left\vert
S_{0}\right\vert B_{\pm}}{\hbar\left(  \omega\pm\omega_{g}\right)  } \label{6}%
\end{equation}
Using Eq. (\ref{6}) to express the free current as well as the magnetization
current in terms of the electric field, and using Eq. (\ref{5}), the following
linear dispersion relation is obtained for the CPEM modes
\begin{equation}
n_{R}^{2}=1-\frac{\omega_{pe}^{2}}{\omega\left(  \omega\pm\omega_{c}\right)
}-\frac{g^{2}\omega_{pe}^{2}k^{2}\left\vert S_{0}\right\vert }{4\omega
^{2}m_{e}\left(  \omega\pm\omega_{g}\right)  }, \label{7a}%
\end{equation}
which can be rewritten as
\begin{equation}
n_{R}^{2}\left(  1+\frac{\omega_{\mu}}{\omega\pm\omega_{g}}\right)
=1-\frac{\omega_{pe}^{2}}{\omega\left(  \omega\pm\omega_{c}\right)  },
\label{7}%
\end{equation}
where $n_{R}\equiv ck/\omega$ is the refractive index, and where the upper and
lower sign respectively stand for the left-hand circularly polarized (LCP) and
right-hand circularly polarized (RCP) waves. Also, $\omega_{\mu}=g^{2}%
\hbar/8m_{e}\lambda_{e}^{2}$ is a frequency which involves the spin correction
due to plasma magnetization current and $\lambda_{e}\equiv c/\omega_{pe}$ is
the electron skin-depth (inertial length scale). Moreover, $\omega_{pe}%
\equiv\sqrt{n_{0}e^{2}/\varepsilon_{0}m_{e}}$, $\omega_{c}=eB_{0}/m_{e}\ $and
$\omega_{g}=(g/2)\omega_{c}$ \ are respectively the electron plasma, cyclotron
and the spin-precession frequency. In absence of the spin-magnetization, the
well-known classical dispersion relation, namely $\omega^{2}=c^{2}k^{2}%
+\omega\omega_{pe}^{2}/\left(  \omega\pm\omega_{c}\right)  $ is recovered. \ A
few comments are in order. The first and the second term in the right-hand
side of Eq. (\ref{7a}) appear due to the displacement current and the free
electron current. The term involving $\omega_{\mu}$ appears even in absence of
the external magnetic field, since the spin perturbation is due to the wave
magnetic field not the constant field $B_{0}$ which does not provide any
magnetic dipole force. Note, however, that an unperturbed spin state with
$\left\vert \mathbf{S}_{0}\right\vert =\hbar/2$, that has been used in the
derivation, typically requires that the external magnetic field is strong.
Thus, inclusion of the electron spin perturbation leads to a modification of
the dispersion relation for transverse \ plasma oscillations. This
modification is clearly substantial when $\omega<\omega_{g}\lesssim$
$\omega_{\mu},$ i.e., when $\hbar\omega_{c}\gtrsim m_{e}c^{2}\ $for
$\omega_{pe}\lesssim\sqrt{2}\omega_{c},$ where $c$ is the speed of light in
vacuum. This corresponds to a regime of very strong magnetic field in which
the external field strength approaches or exceeds the quantum critical
magnetic field, i.e., $B_{0}\gtrsim$ $B_{Q}\equiv4.4138\times10^{9}T.$ In such
a situation relativistic effects might be important. On the other hand, for
the nonrelativistic motion of electrons we have $\hbar\omega_{c}<m_{e}c^{2},$
i.e., $B_{0}<$ $B_{Q}$ for $\omega_{pe}>\sqrt{2}$ $\omega_{c}$. In this case,
the density regime in which the magnetic field is `nonquantizing' and does not
affect the thermodynamic properties of the electron gas, is $n_{0}\gg
n_{c}\simeq10^{32}$m$^{-3}$ and the temperature $T_{e}\gtrsim T_{B}\simeq
\hbar\omega_{c}/k_{B},$ where $k_{B}$ is the Boltzmann constant. Thus, in the
strong magnetic field and highly dense plasmas, the electron spin effect can
no longer be neglected, rather it modifies the wave dispersion leading to new eigenmodes.

Inspecting now the term $\propto\hbar$ in Eq. (\ref{7a}), we note that%

\begin{equation}
\frac{\hbar k^{2}}{m_{e}\omega}=\left(  \frac{\hbar\omega_{c}}{m_{e}c^{2}%
}\right)  \left(  \frac{c^{2}k^{2}}{\omega^{2}}\right)  \left(  \frac{\omega
}{\omega_{c}}\right)  . \label{8}%
\end{equation}
So, the spin current can be much larger than the classical free current when
$\left\vert J_{M\pm}/J_{C\pm}\right\vert \sim\hbar k^{2}/m_{e}\omega \gg 1.$This
basically holds when either (i) $\hbar\omega_{c}\gtrsim m_{e}c^{2},$
$\omega<\omega_{c},ck$ or (ii) $\hbar\omega_{c}<m_{e}c^{2},$ $\omega
<\omega_{c}\ $and $\omega \ll ck$ is satisfied. \ The case of $\omega
>\omega_{g}>\omega_{c}$ is rather less important, as it does not give rise to
wave propagation ($n_{R}^{2}<0$). Also, in the very lf regime $\omega\ll
\omega_{c},$ the ion motion can be of importance, and we will therefore not
consider that case.\ Thus, one important mode could be the RCP lf
($\omega<\omega_{c}$) EM waves (whistlers). Let us now see how the dispersion
relation reduces when either of the two cases is considered. In the limit of
$\left\vert J_{M\pm}/J_{C\pm}\right\vert \sim\hbar k^{2}/m_{e}\omega \gg 1$, the
dispersion equation (\ref{7}) reduces to%

\begin{equation}
\left(  \omega\pm\omega_{g}\right)  \left(  1-\omega^{2}/c^{2}k^{2}\right)
+\omega_{\mu}=0, \label{9}%
\end{equation}
from which one finds for $\omega\ll ck$ a purely spin-modified frequency
$\omega_{1}\approx\mp\omega_{g}-\omega_{\mu}$. Also, if $\omega_{\mu}\ll ck$
and $\omega_{\mu}\ll\omega_{g},$ we have $\omega_{2}\approx\mp\omega_{g}$ and
$\omega_{3}^{2}\approx c^{2}k^{2}.$ The frequencies $\omega_{1,2}$ may
correspond to the spin waves in ferromagnets under certain conditions
\cite{Dyson}. Figure 1 displays the modes for RCP waves obtained as numerical
solutions of the dispersion equation. Evidently, there exist two eigenmodes
apart from a hf ($\omega>\omega_{g})$ one, namely a mode close to the
electron-cyclotron or spin-precession frequency, and the other one is the lf
mode below the cyclotron frequency. In contrast to the hf modes, the spin
modified lf modes propagate with the frequency below that in the classical
case. On the other hand, in the very hf regime ($\omega\gg\omega_{g}),$
$n_{R}^{2}$ $\approx1>0,$ and so we have $\omega^{2}=c^{2}k^{2}$, which is the
dispersion relation of a EM wave in vacuum. Evidently, since $n_{R}^{2}<0$ for
$\omega>\omega_{g},$ there must exist an intermediate frequency $\omega
(>\omega_{g})$ at which the solution for $n_{R}^{2}$ must pass through a zero
value, and becomes positive again. \ Thus, the cut-off frequencies for which
$n_{R}^{2}=0$ are obtained as%

\[
\omega_{R,L}=\frac{1}{2}\left(  \mp\omega_{c}+\sqrt{\omega_{c}^{2}%
+4\omega_{pe}^{2}}\right)  ,
\]
where $\mp$ stand for RCP and LCP waves respectively. Clearly, the RCP waves
have lower cut-off frequency than the LCP modes. On the other hand, the
resonances for the RCP waves ( $n_{R}^{2}\rightarrow\infty)$ associated with
both the orbital and the spin-gyration, occur (LCP mode\ has no resonance) as
either $\omega\rightarrow\omega_{c}$ or $\omega\rightarrow\omega_{g},$ i.e.,
when the angular frequency of the wave electric field matches either due to
electron cyclotron motion (cyclotron resonance) or due to the intrinsic spin
of electrons (ESR). At the resonance, the transverse field associated with the
RCP wave rotates at the same velocity as electrons gyrate around $B_{0}.$ The
electrons thus experiences a continuous acceleration from the wave electric
field, which tends to increase their perpendicular energy. Therefore, it is
not surprising that RCP waves propagating along the the external magnetic
field and oscillating at the cyclotron frequency or spin-precession frequency
are absorbed \ by electrons. This may be the consequence to the recently
developed experiment based on microwave absorption and the ESR to be
successfully used for plasma diagnostics \cite{EPR}. On the other hand, since
the spin effect is appreciable in the strongly magnetized dense plasmas, the
ESR could well be relevant for the coherent EM radiation in pulsar
magnetosphere or magnetized white dwarfs. \begin{figure}[ptb]
\begin{center}
\includegraphics[height=3in,width=4in]
{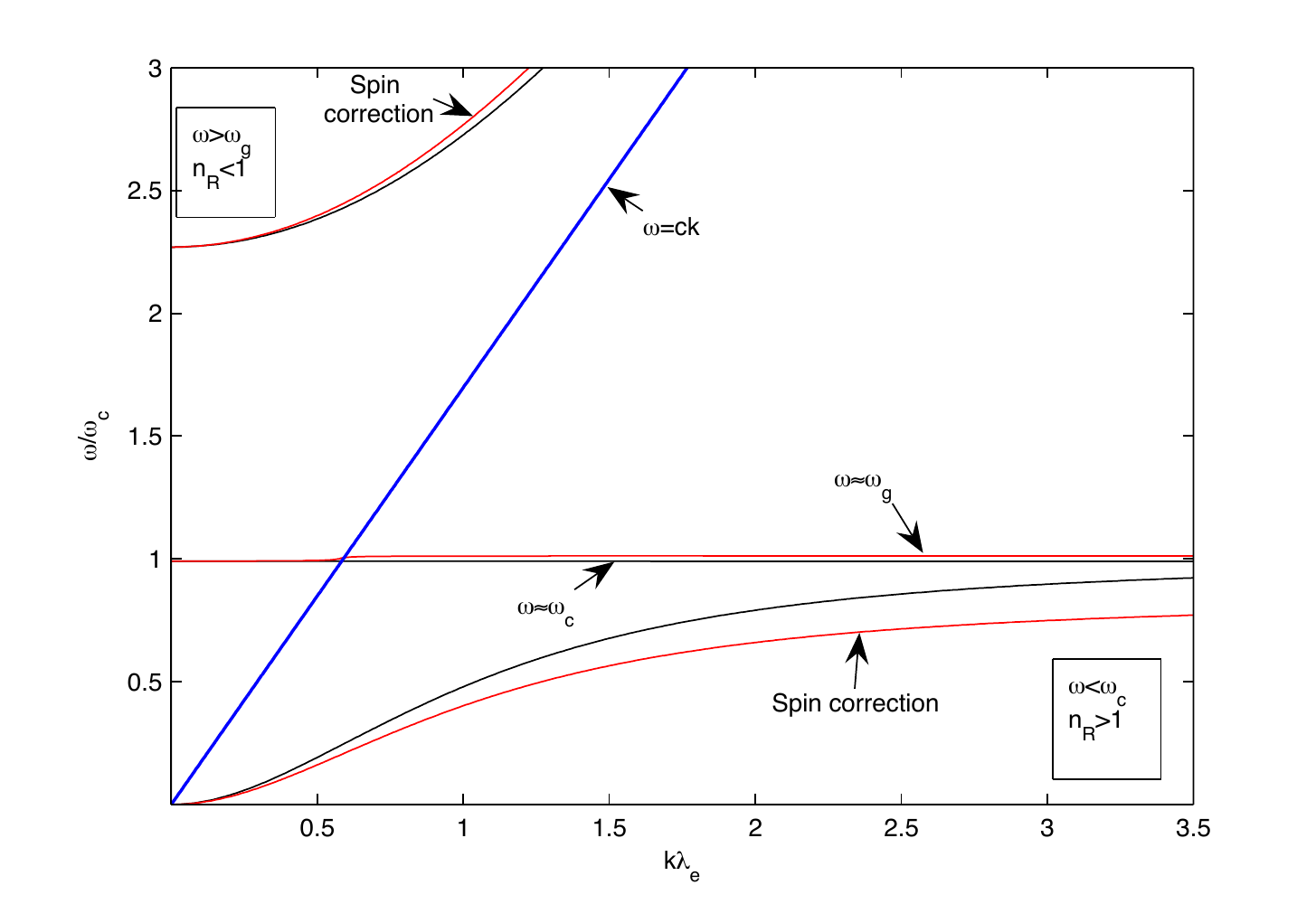}
\end{center}
\caption{Different eigen modes for RCP waves obtained as numerical solutions
of the dispersion equation (\ref{7a}) are shown with respect to the normalized
wave number and frequency for $B_{0}=5\times10^{8}\, \mathrm{T} < B_Q$, $n_{0}%
=7\times10^{36}\, \mathrm{m}^{-3} \gg n_{c}.$}%
\end{figure}\begin{figure}[ptb]
\begin{center}
\includegraphics[height=3in,width=4in]
{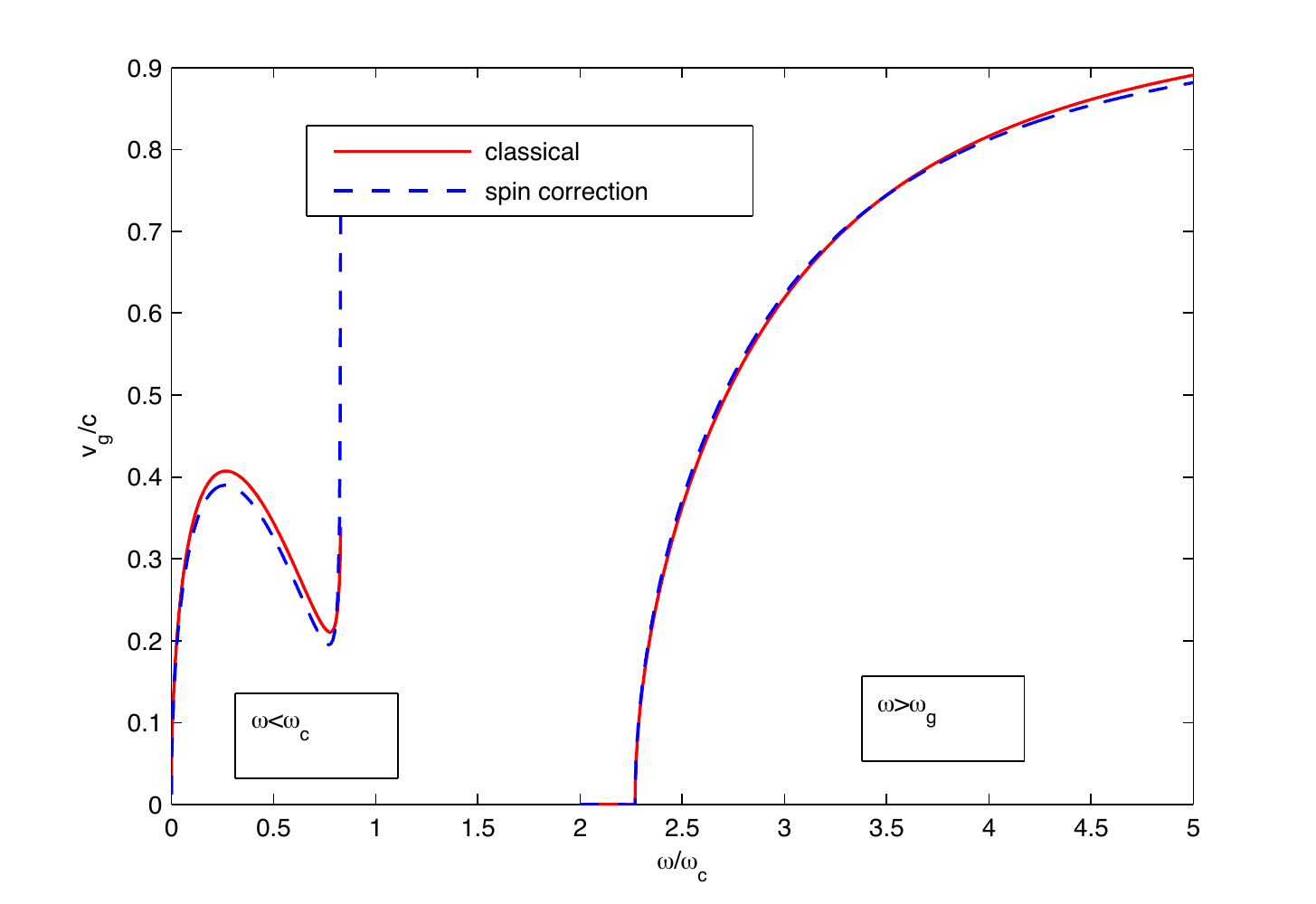}
\end{center}
\caption{The normalized group speed given by Eq. (\ref{group}) is plotted as a
function of the normalized wave frequency for the RCP waves with the same set
of parameters as in Fig. 1. }%
\end{figure}Let us now see how the group speed $\left(  v_{g}\equiv
d\omega/dk\right)  $ of the CP waves is modified with the spin correction. We
obtain from the dispersion relation [Eq. (\ref{7a})]%

\begin{equation}
v_{g}=\Lambda/\left(  2\omega/\omega_{pe}^{2}+\Gamma\right)  , \label{group}%
\end{equation}
where\ $\Lambda$ and $\Gamma$ are given by%

\begin{equation}
\Lambda=\frac{2c^{2}k}{\omega_{pe}^{2}}+\frac{g^{2}k\left\vert S_{0}%
\right\vert }{2m_{e}\left(  \omega\pm\omega_{g}\right)  },\text{ }\Gamma
=\frac{\omega_{c}}{\left(  \omega\pm\omega_{c}\right)  ^{2}}+\frac{g^{2}%
k^{2}\left\vert S_{0}\right\vert }{4m_{e}\left(  \omega\pm\omega_{g}\right)
^{2}}.
\end{equation}
Clearly, the lf $\left(  \omega<\omega_{c}\right)  $ component of \ a pulse
(whistlers) propagates more slowly than the hf $\left(  \omega>\omega
_{g}\right)  $ components as is evident from Fig. 2. \ It follows that by the
time a pulse returns to a ground level it has been stretched out temporarily,
because the hf component of the pulse arrives slightly before the lf
components. This also accounts for the characteristic whistling down-effect
observed at ground level. Moreover, the group speed $v_{g}$ of the whistlers
increases in the frequency regime $0<\omega<\omega_{c}/2,$ and decreases in
the other subinterval $\omega_{c}/2\lesssim\omega<\omega_{c}$ before it
reaches the maximum nearer the resonance point. However, the group speed of
the hf modes approaches the speed of light as $\omega$ $\left(  >\omega
_{g}\right)  $ increases and gets saturated at large $\omega.$ From Fig. 2 we
also note that the spin force reduces the group speed in strongly magnetized
dense plasmas.

\section{Summary and discussion}

To summarize, \ the dispersion relation for the propagation of hf CPEM waves
is obtained in a magnetized spin plasma. The electron spin modifies the plasma
current density and thereby introduces a correction term in the dispersion
relation, which, in turn, gives rise a new\ CP hf mode. The spin effects are
seen to be substantial in the very strong magnetic field $\left(  B_{0}\gtrsim
B_{Q}\equiv4.4138\times10^{9}T\right)  $ and highly dense plasmas $\left(
n_{0}\gg n_{c}\simeq10^{32}m^{-3}\right)  $ where the relativistic effects
might be important. However, in nonrelativistic plasmas, the spin of electrons
can also be important in the case $B_{0}<B_{Q}$ together with $n_{0}\gg
n_{c}.$ In particular, when the spin current dominates over the classical free
current the RCP EM waves resonantly interact with the electrons only at the
spin-precession frequency. Such resonance should be helpful for particle
acceleration in the coherent radio emission of the pulsar magnetosphere or
magnetized white dwarfs. The study of the spin modified modes might also be
important at least from the diagnostic points of view, since the observation
of the propagation characteristics of the wave modes may be used in order to
determine the physical parameters in plasmas \cite{Diagnostics}. Lastly, the
electron spin-resonance (ESR) could be an important consequence to the
recently developed experiment for plasma diagnostics in the microwave region,
if the conditions favor \cite{EPR}.

\acknowledgments

A. P. M. gratefully acknowledges support from the Kempe Foundations, Sweden. MM was supported by the Swedish Research Council Contract \# 2007-4422 and the European Research Council Contract \# 204059-QPQV

\end{document}